\documentclass[conference]{IEEEtran}

\newif\ifincluded
\includedfalse

\ifincluded
\fi

\usepackage[flushleft]{threeparttable}

\usepackage{listings}

\ifCLASSOPTIONcompsoc
  \usepackage[nocompress]{cite}
\else
  \usepackage{cite}
\fi

\usepackage{amsfonts}  

\usepackage{amsmath}  

\usepackage{multirow}

\usepackage{hyperref}

\usepackage{graphicx}

\begin{document}
\bstctlcite{IEEEexample:BSTcontrol}

%
\title{Anonymous State Pinning for Private Blockchains}


\author{
    \IEEEauthorblockN{Peter Robinson}
    \IEEEauthorblockA{Protocol Engineering Group and Systems, ConsenSys\\
    peter.robinson@consensys.net}
    \IEEEauthorblockA{School of Information Technology and Electrical\\
    Engineering, University of Queensland, Australia\\
    peter.robinson@uqconnect.edu.au}
\and
    \IEEEauthorblockN{John Brainard}
    \IEEEauthorblockA{Protocol Engineering Group and Systems, ConsenSys\\
    john.brainard@consensys.net}
}

\maketitle

\thispagestyle{plain}
\pagestyle{plain}


\begin{abstract}

Public blockchains such as Ethereum and Bitcoin provide transparency and accountability, and have strong non-repudiation properties, but fall far short of enterprise privacy requirements for business processes. Consequently consortiums are exploring private blockchains to keep their membership and transactions private. However, private blockchains do not provide adequate protection against potential collusion by consortium members to revert the state of the blockchain. To countenance this, the private blockchain state may be ``pinned" to a tamper resistant public blockchain. Existing solutions offering pinning to the public blockchain would reveal the transaction rate of the private blockchain, and do not provide a mechanism to contest the validity of a pin. Moreover, they require that all transactions and members of the private blockchain be revealed. These challenges are hampering the wider adoption of private blockchain technology. We describe the primary author's `Anonymous State Pinning approach', which overcomes these limitations and present a security proof to demonstrate pins can be challenged without compromising these properties. We perform a gas cost analysis of the implementation to estimate the operating cost of this technology, which shows that pinning a private blockchain at the rate of one pin per hour would cost US\$508 per year. A hierarchical pinning approach is proposed which would allow many private blockchains to pin to a management blockchain which would then pin to Ethereum MainNet. This approach saves money, but at the cost of increased finality times.

\end{abstract}


%
\IEEEpeerreviewmaketitle

\section{Introduction}
In this paper we focus on private consortium blockchains, and specifically on the ``Anonymous State Pinning'' approach previously proposed by the primary author, to make private blockchains more secure and less vulnerable to collusion by members attempting to revert the blockchain state \cite{robinson2018a}. We propose a delegate blockchain architecture which allows the security properties of Ethereum MainNet to be leveraged whilst not impacting the performance of MainNet or incurring significant transaction costs. 

We organised the paper as follows to provide context for the proposed approach: the \textit{Background} section has a brief introduction to blockchain, and then describes Ethereum, the platform on which the methodology is built. We introduce the concept of private blockchains and the enterprise version of Ethereum, explain what `state pinning' is and the importance of having `finality' for blocks so that they can be added to the blockchain. The next section, \textit{Related Works}, reviews alternative existing techniques for pinning, showing that they are not appropriate for private blockchains. The remaining sections describe the approach in detail, how pins can be contested, and how participants can be added and removed  \cite{robinson2018a}.  We demonstrate, via a formal security proof, that selected pieces of information to challenge a pin does not reveal previous or future state pins, hence preventing malicious activity to hamper the posting of valid pins, and keeping the pinning rate private. We analyse the performance of the author's Solidity code implementation, which is available on Github \cite{pinning_github}. 

\section{Background}
\subsection{Blockchain}
Satoshi Nakamoto \cite{nakamoto2008} created Bitcoin, the first blockchain platform, in 2008. Blockchains are cryptographically linked lists of blocks of transactions. Each transaction changes the state of accounts. Replaying the transactions in all blocks in order yields the current state of all accounts, known as the distributed ledger. Blockchain miners agree on which transactions should be included in the next block using a consensus algorithm. 

Transactions are combined into a block by message digesting the transaction and including that digest in a Merkle Tree. The Merkle root of all the transactions to be included in the block is included in the block header. Linkage between blocks is achieved by including the message digest of the previous block header in the current block header. Block headers also include other information such as block number and consensus algorithm values. This message digest of a block header, known as a ``Block Hash" succinctly captures all of the transactions from the genesis block to the current block, thus capturing the state of the distributed ledger.

Users are identified as account holders using ECC key pairs. Bitcoin and Ethereum have used the secp256k1 curve defined in Certicom Research's, ``Standards for Efficient Cryptography - SEC 2: Recommended Elliptic Curve Domain Parameters" \cite{certicom2010}. User's account numbers are 160 bit message digests of the user's public key. Thus, a transaction signed with an account holder's private key can be verified using the account number by first deriving the public key from the transaction signature and then message digesting the public key and comparing it with the account number.

\subsection{Ethereum}
Ethereum \cite{wood2016a} builds on the Bitcoin platform, allowing users to upload and execute computer programs known as Smart Contracts. Ethereum Smart Contracts can be written in a variety of Turing complete languages, the most popular being Solidity \cite{solidity}. Code is compiled into a bytecode representation. The bytecode can then be deployed using a contract creation transaction. Contracts have a special function called \texttt{init} which only runs when the contract creation transaction is being processed. This function is used to initialize memory and call other contract code. Miners execute the bytecode inside virtual machines. At present, each miner must execute all transactions for all contracts and hold the current value of all the memory associated with all of the contracts. The Ethereum community is actively working on methodologies to scale the Ethereum network by sharding the blockchain \cite{ethereum-sharding}.

Ethereum transactions update the state of the distributed ledger. They fall into three categories: contract creation, calling a function on a contract, and transferring Ether. Ether transfer transactions moves Ether from the user's account to another account. Contract creation transactions put code into the distributed ledger and call the constructor of the contract code, setting the contract data's initial state. Function call transactions call a function on a contract and result in updated state. All types of transactions must be signed by a private key corresponding to an account and include a nonce value which prevents replay attacks. In addition to Ethereum transactions, ``View" function calls can be executed on the Smart Contract code. These View function calls return a value and do not update the state of the Smart Contract.

Executing code and accessing resources, such as memory, costs certain amounts of ``Gas". The ``Gas Cost'' of executing code is closely tied to the real world cost of executing each type of instruction. The current ``Gas Price'' is set for each block in terms of Ether by the miner who mines the block. Accounts instigating transactions specify the gas price they are prepared to pay for their transaction and specify the maximum amount of gas a transaction can use known as ``Start gas". This commits an account holder to paying up to a certain amount of Ether for the transaction. Any unused gas is returned to the account holder at the end of the transaction. Miners reject transactions which run out of gas prior to completing execution.

In the Ethereum public network, ``MainNet", all contract code and data are readable by any user of any node which connects to the network. Smart Contracts on Ethereum MainNet can only perform permissioning in contract code, limiting which accounts can update the state of a contract. However, there is no mechanism to limit which users can read contract code and data. 

The value proposition of Ethereum is that it allows untrusted parties to use Smart Contracts hosted on a public, distributed, highly available, secure platform. 

\subsection{Private Blockchains and Enterprise Ethereum}
Private blockchains are blockchain networks which are established between nodes operated by enterprises \cite{robinson2018a}. Only permissioned nodes belonging to participating enterprises are allowed to join the private blockchain's peer-to-peer network and only permissioned accounts belonging to participating enterprises are allowed to submit transactions to the nodes. These blockchains provide the privacy and permissioning required by enterprises \cite{enteth20}. 

The need for security and permissioning features over and above what is available in standard Ethereum \cite{enteth20} has led to a range of platforms being developed. J.P. Morgan developed Quorum \cite{quorum-source}, a fork of the Golang Ethereum implementation called, geth  \cite{geth-github}. Parity Technologies added a Private Transactions feature to their existing Ethereum client. ConsenSys's Protocol Engineering Group, PegaSys created Pantheon \cite{pantheon-github}, an Ethereum MainNet compatible client which aims to meet the permissioning and privacy requirements of the Enterprise Ethereum Client Specification \cite{enteth20}. Hyperledger Fabric \cite{androulaki2018} is a distributed ledger platform originally created by IBM and now hosted by The Linux Foundation. This platform directly competes with Quorum and Parity, offering privacy and permissioning features. Whereas Quorum and Parity offer private smart contracts which operate on or in conjunction with a permissionless blockchain, Hyperledger Fabric offers the ability to host one or more smart contracts on a private blockchain called a ``channel". Hyperledger Fabric allows multiple channels to be operated on the one network, thus allowing for multiple sets of private contracts between different sets of participants to operate on the one network.

The strong privacy and permissioning features required by enterprises are poorly served by the permissionless Ethereum MainNet  \cite{wood2016a} which provides strong authenticity and non-repudiation properties, but no privacy or confidentiality \cite{xu2017}. However, these strong non-repudiation properties mean that Ethereum MainNet is the ideal location for securely storing data for which the authenticity and integrity of the data is paramount. 

\subsection{State Pinning}
State Pinning is defined as putting the Block Hash of one blockchain into another blockchain. For instance, including the Block Hash of a Private Blockchain in Ethereum MainNet. As the Block Hash from the Private Blockchain is included in Ethereum MainNet at a particular block number, it indicates that the state of the Private Blockchain can be represented by that Block Hash at that time. A majority of participants of the Private Blockchain could collude to alter the historical state of the chain \cite{kaleido-relay}. State Pinning allows minority participants of the chain to prove to governmental regulators and others that the state of the chain has been altered, by showing that the correct state matches the pinned Block Hash.

When the state of a consortium blockchain is pinned it is important to maintain the privacy of blockchain participants by not revealing the participants of the blockchain or the blockchain transaction rate. Not disclosing the transaction rate of a private blockchain is important as attackers may be able to infer activity based on this.

\subsection{Finality}
A block is deemed final when it can no longer be changed. In some consensus algorithms, such as PoW, finality is probabilistic, where as more blocks are added to the end of the blockchain, older blocks are less likely to be reordered. Consensus algorithms such as Istanbul Fault Byzantine Tolerant (IBFT) \cite{ibft} give ``instant" finality, where once a transaction has been included in a block minted by a validator, it can not be changed.

\section{Related Works}
This section outlines existing techniques for pinning the state of one blockchain to another blockchain.

\subsection{Merge Mining}
Merge Mining \cite{namecoin2015}\cite{merged-mining-2011}
\cite{merged-mining2014} 
is a technique in which the Block Hash of a low hashing power public blockchain, such as NameCoin, is included in a more secure higher hashing power blockchain, such as Bitcoin. In this scenario, the Bitcoin miners must validate NameCoin transactions prior to including the Block Hash in a transaction on the Bitcoin network. The mined transaction can then be included in both the Bitcoin and NameCoin blockchain. Merged Mining relies on both blockchains using the same consensus algorithm, and assumes that all transactions can be viewed by both blockchains. As such, this technique is not usable in a private blockchain scenario where transactions must not be revealed outside the blockchain.

\subsection{Tethered Permissioned Private Chains}
After discussions with the primary author, the Kaledio team developed Tethered Permissioned Private Chains \cite{kaleido}\cite{kaleido-relay} to reduce the risk of state reversion occurring by posting the state of the blockchain onto Ethereum MainNet. In Kaleido's system, an entity similar to the proposed Quiet Guardian \cite{robinson2018a} may be used to submit Pins on behalf of blockchain participants, and has to reveal itself. Moreover, the rate of transactions on the consortium chain is revealed on Ethereum MainNet due to the number of posted Pins. Additionally, the solution lacks a method of contesting a Pin posted to Ethereum MainNet.

\subsection{Polkadot}
In Polkadot \cite{polkadot2016} Para-Chain blocks are sealed on the Relay Chain after transactions and zero knowledge non-interactive proofs proving the transactions result in valid state changes are validated. The block sealing involves including the block header hash in the Relay Chain. This system requires Validators which are randomly allocated to the Para-Chain to be able to view transactions on the Para-Chain and then posting to the Relay Chain. This requirement for randomly allocated Validators to view information on the Para-Chain is incompatible with the concepts of private blockchains which need to restrict the list of participants. Additionally, having Validators post Block Hashes will reveal the Validators to all members of the blockchain system. 

\subsection{BTC Relay}
BTC Relay \cite{btc-relay} is a method for allowing users of Ethereum to confirm Bitcoin transactions. Relayers are compensated for posting Bitcoin block headers to a Smart Contract on Ethereum. Bitcoin transactions are confirmed by users submitting Merkle proofs showing that a transaction belonged to a certain block. From a blockchain state pinning perspective, BTC Relay pins the state of the Bitcoin blockchain to Ethereum.

BTC relay relies on PoW mining difficulty for its security. Multiple active Relay nodes must be prepared to post the block header for each block. In this way, if one Relay node posts a block header of a fork of the chain, other Relay nodes can post the block header of the longest chain. Transactions can only be validated if the block header they relate to is on the longest chain and if at least six block headers have been posted on top of the block header that the transaction relates to \cite{btc-relay-source}. As attackers can not produce a longer chain than the main Bitcoin blockchain due to the mining difficulty, they are unable to confirm transactions based on a malicious fork.

PoW is not an appropriate consensus algorithm for private blockchains as organisations do not wish to allocate resources to mining of blocks \cite{enteth20}. Given the reliance of BTC relay on PoW mining difficulty, it is inappropriate for pinning private blockchains.

\subsection{Summary}
None of the existing state pinning approaches are appropriate for pinning the state of private blockchains. In particular, none of the approaches allows pinning of private state, without revealing transactions, whilst keeping the participants of the private blockchain secret, and allowing those participants to challenge the pinned state. 

Private blockchain participants need to keep the rate of transactions secret. If state pins are related to transactions or blocks of transactions, then the posted Pins can betray the rate of transactions on the private blockchain. None of the existing techniques provided any method of obscuring or hiding the transaction rate of the private blockchain.

\section{Architecture}
\subsection{Introduction}
To overcome the limitations of existing solutions, a methodology was proposed by the primary author \cite{robinson2018a} which allows private blockchain state to be anonymously pinned to an Ethereum blockchain, such as Ethereum MainNet. The chain being pinned to is known as the ``Management Chain". The private blockchain state is represented as the Block Hash for a block on the private blockchain to be pinned to the Management Chain. The Block Hash of the block is known as a ``Pin". Pins are posted into a key-value map, in which the key is specially crafted, and the value is the Pin. The key values are derived using a technique inspired by the ideas of Code Division Multiple Access \cite{elbert1987} in which multiple sets of information can be overlaid on the same channel without interfering with each other. The technique is similar to the use of Pseudo Random Functions for identifying friend or foe and for ``storageless" distribution of random numbers  \cite{goldreich1985} and the use of salted hashes for secure communications \cite{kent2011} in that random looking numbers are derived from a combination of public and secret values.

\subsection{Masked and Unmasked Participants}
For each private blockchain, there are masked and unmasked participants. Unmasked Participants have their Ethereum Addresses listed as being members of the private blockchain. Being unmasked allows the participant to vote to add or remove other participants, and to contest Pins.

Masked Participants are participants that are listed against a private blockchain in a way that observers can not determine their identity. These participants are represented as a salted hash of their Ethereum Address. Each masked participant keeps their secret salt value off-chain. A masked participant may choose to unmask themselves, for example to vote to contest a Pin. To unmask themselves, they present their secret salt to the contract. This combined with their sending address is used to create the calculated salted hash. If this calculated value matches their masked participant value then they become an unmasked participant.

\subsection{Map Key Calculation}
Pinning values are put into a key-value map. All participants of a private blockchain agree on a Private Blockchain Identifier (PBI), a Private Blockchain Secret and a Pseudo Random Function (PRF) algorithm. The PBI is a public value used to specify the private blockchain. The Private Blockchain Secret seeds the PRF. A new 256 bit value is generated by the PRF each time an uncontested Pin is posted. The key in the map is calculated using the equation shown below.

\begin{equation}
MapKey\textsubscript{t} = KECCAK\textnormal{-}256( PBI, Pin\textsubscript{t-1}, PRF(t))
\end{equation}
Where the initial Pin value, Pin\textsubscript{-1} is zero. 

\subsection{Contesting a Pin}
Masked and unmasked participants of a private blockchain observe the pinning map at the MapKey\textsubscript{t} address corresponding to the next Pin, waiting for the next Pin to be posted to that entry in the map. When the Pin value is posted, they check that the posted Pin matches their understanding of the most recent Block Hash of the private blockchain. If the values do not match, then participants should contest the Pin. To contest the Pin, they submit to the contract: MapKey\textsubscript{t-1}, PRF(t), and the PBI.

Submitting the previous value of the MapKey allows the contract to fetch from its own storage the value of the previous Pin, Pin\textsubscript{t-1}. The contract can then calculate the MapKey of the contested Pin, MapKey\textsubscript{t}, by combining Pin\textsubscript{t-1}, PRF(t) and the PBI using the equation above. Given the submitter of the transaction knows the PRF(t) which combined with the PBI links the previous MapKey, MapKey\textsubscript{t-1}, and the calculated MapKey, MapKey\textsubscript{t}, it implies that both of the MapKeys correspond to Pins for the private blockchain denoted by PBI. The further implication of knowing PRF(t) is that the transaction submitter has access to the Private Blockchain Secret, which implies that they are a member of the private blockchain.

Once a Pin is marked as contested, the unmasked participants of the private blockchain must then vote on the validity of the posted Pin. At this point, masked participants will need to unmask themselves to vote. This could be viewed as analogous to a situation in which companies that have a traditional private written agreement are in dispute. Under normal circumstances the companies may be able to keep their agreement private. However, in case of dispute, the companies need to make public their agreement and go to the courts to resolve the matter. Similarly, masked participants need to unmask themselves if they wish to vote on a disputed Pin. 

Pins can only contested for a fixed period known as the ``Pin Dispute Period". After this period has passed, Pins can be viewed as being final.

If a Pin has been successfully contested, the Pin value is replaced with the value \texttt{0xFFFFFFFF} and all participants use the next PRF value PRF(t+1) to calculate a new MapKey value, MapKey\textsubscript{t}\textsuperscript{*}.

\subsection{Hiding Participants}
Pins are posted to the Management Chain using Ethereum transactions. As all Ethereum transactions are signed, this means that the act of posting a Pin reveals that the entity posting the Pin is a participant of the private blockchain. To hide the true participants of the private blockchain, the blockchain transactions should be encrypted using a key known to blockchain nodes belonging to true participants, but not to a special participant called a Quiet Guardian. As the Quiet Guardian's node does not have access to the shared key, it can only see transaction cipher text, and see the block header information such as the Block Hash. As such, the Quiet Guardian can post the Block Hash pins to the Management Chain, thus earning the ``Guardian" part of its name, but can not submit transactions to the private blockchain, thus earning the ``Quiet" part of the name. The Quiet Guardian could be the only unmasked participant of a private blockchain. Having a Quiet Guardian as the lone unmasked participant of a private blockchain allows the other blockchain participants to remain anonymous, assuming no dispute. 

\subsection{Hiding Pinning Rate}
Multiple private blockchains could be managed from the one Anonymous Pinning smart contract. This would be advantageous because it would mean that the Pin values for the different private blockchains could be intermixed in the one pinning map. This would hide the rate of pinning for any one blockchain, assuming the same account was used to post Pins for all blockchains, as attackers could not discern which blockchain each Pin was for.

\subsection{Hierarchical Pinning}
Pins are posted to the Management Chain using Ethereum transactions. If the Management Chain is Ethereum MainNet, then each of the Ethereum transactions will cost money to execute (see section \ref{sec:gas} for an analysis of how much). If many private blockchains were posting pins to Ethereum MainNet, this could cause significant network load as happened due to the Crypto Kitties game \cite{crypto-kitties} and other games which use Ethereum MainNet \cite{crypto-kitties-other}. 

\begin{figure}
  \includegraphics[width=\linewidth]{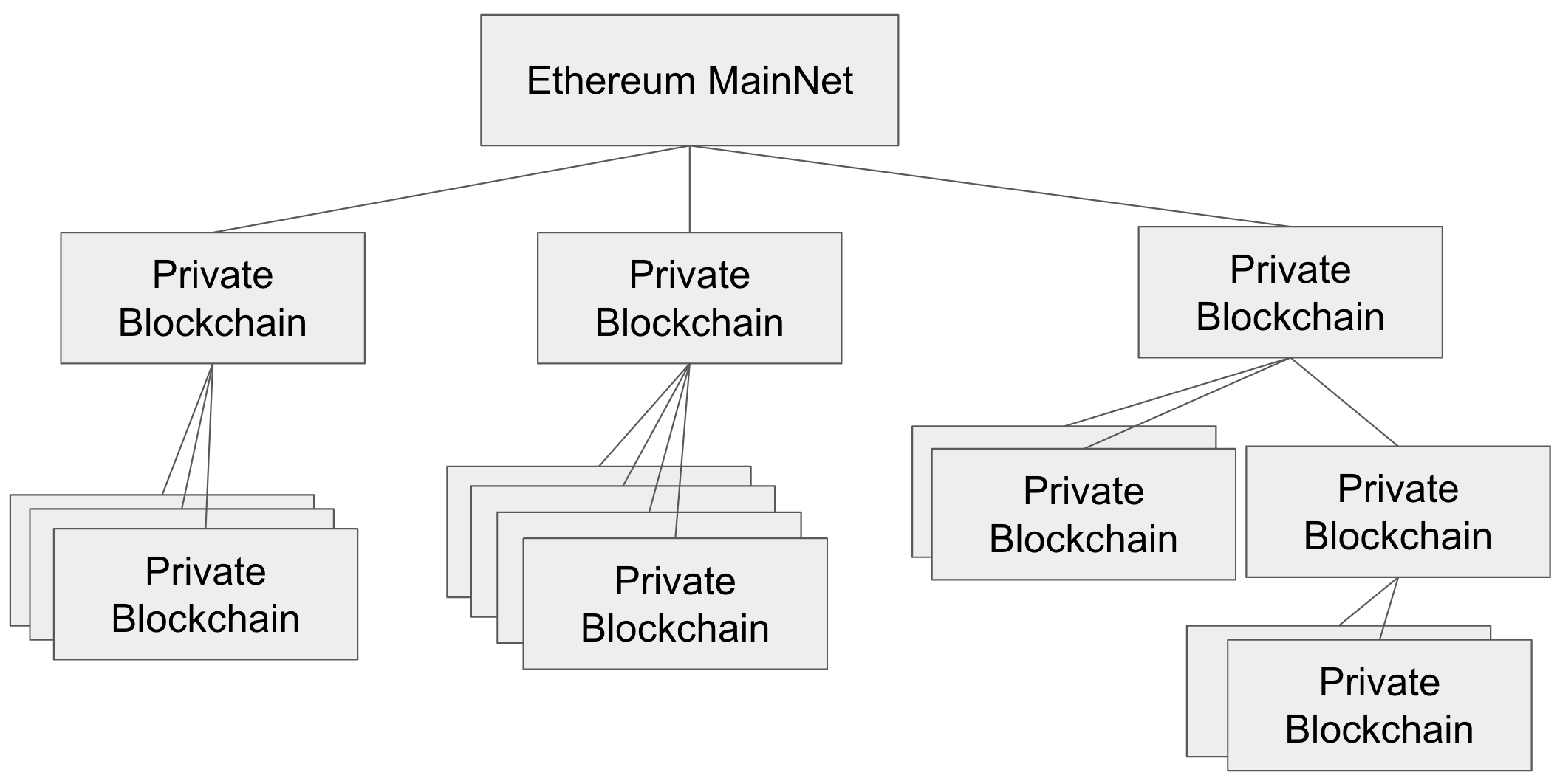}
  \caption{Hierarchical Pinning}
  \label{fig:hierarchical}
\end{figure}

To over come the problems of cost of transactions and the possibility of congestion on Ethereum MainNet, a hierarchical pinning model is proposed. With this model, many private blockchains could treat another private blockchain as a Management Chain posting Pins to it. This private blockchain could in turn post Pins to another private blockchain or to Ethereum MainNet. This is shown diagrammatically in Figure \ref{fig:hierarchical}. Pinning to a hierarchy of Management Chains in this way means that only a small number of Pins on Ethereum MainNet could be used to secure a large number of private blockchains.

An additional benefit of pinning to a private blockchain is that the chain's permissioning could be set such that only certain nodes could view the blockchain and only certain accounts could submit transactions to the blockchain. 

There are three areas of concern with respect to the hierarchical pinning approach. The first is that participants of a private chain must observe and be ready to challenge Pins being posted at each level of the hierarchy. The second is that in order for participants of the private blockchain to be able to challenge at each level of the hierarchy, then need to be participants of each blockchain at each level of the hierarchy. The third concern is related to the affects on finality. This third issue is more complex and handled in a section in the analysis section of this paper.

\subsection{Implementation Details}
The Anonymous Pinning Smart Contract implementation and test code is available on Github \cite{pinning_github}. Some key design decisions that have been used in the design of the code are discussed in the following paragraphs. To keep the text below consistent with the source code, the term ``Sidechain" is used in the text below to mean a single Private Blockchain.

When the Smart Contract is first deployed, a sidechain entry is created for a ``Management Sidechain". This Management Sidechain's sole purpose is to list the participants who can add a sidechain entry to the contract. This Management Sidechain has a fixed sidechain identifier of \texttt{0x00}. No other sidechains can be created with this value.

The Smart Contract's constructor takes as parameters a ``Voting Algorithm", ``Voting Period", and ``Pin Dispute Period". The Voting Algorithm is the address of a Smart Contract which implements the ``Voting Algorithm Interface". The Voting Period specifies how long participants have to vote before a participant can request the votes be tallied and the vote be actioned. These voting values pertain to the Management Sidechain only. The Pin Dispute Period pertains to all Pins, and hence all sidechains.

The function \texttt{addSidechain} which adds a sidechain entry takes as parameters a Voting Algorithm and Voting Period. These Voting Algorithm and Voting Period values set the per-sidechain voting configuration.

The process of disputing a Pin is to observe the pinning map waiting for a Pin to be posted to the expected MapKey using the \texttt{getPin} function. When a Pin is detected which the participant does not agree with, they should call \texttt{proposeVote} to propose that the Pin be rejected. The other unmasked participants of the sidechain can then vote using the \texttt{vote} function. Once the Voting Period expires, any sidechain participant can request the vote be tallied and actioned if successful using the \texttt{actionVotes} function. The \texttt{actionVotes} function must be called prior to the Pin Dispute Period expiring. The implication of this is that the Pin Dispute Period must be greater than the Voting Period.

\section{Gas Usage Analysis}
\label{sec:gas}
Gas is the fee charged for each instruction executed in Ethereum. Different instructions are charged different amounts of gas, with the fees reflecting the economic cost of executing the instruction. Users specify the price they are prepared to pay for the gas in each of their transactions. Miners preferentially include transactions in blocks which are configured to pay the highest gas price. As such, transactions which are submitted with a higher gas price are more likely to be included in any given block. 

\begin{figure}
  \includegraphics[width=\linewidth]{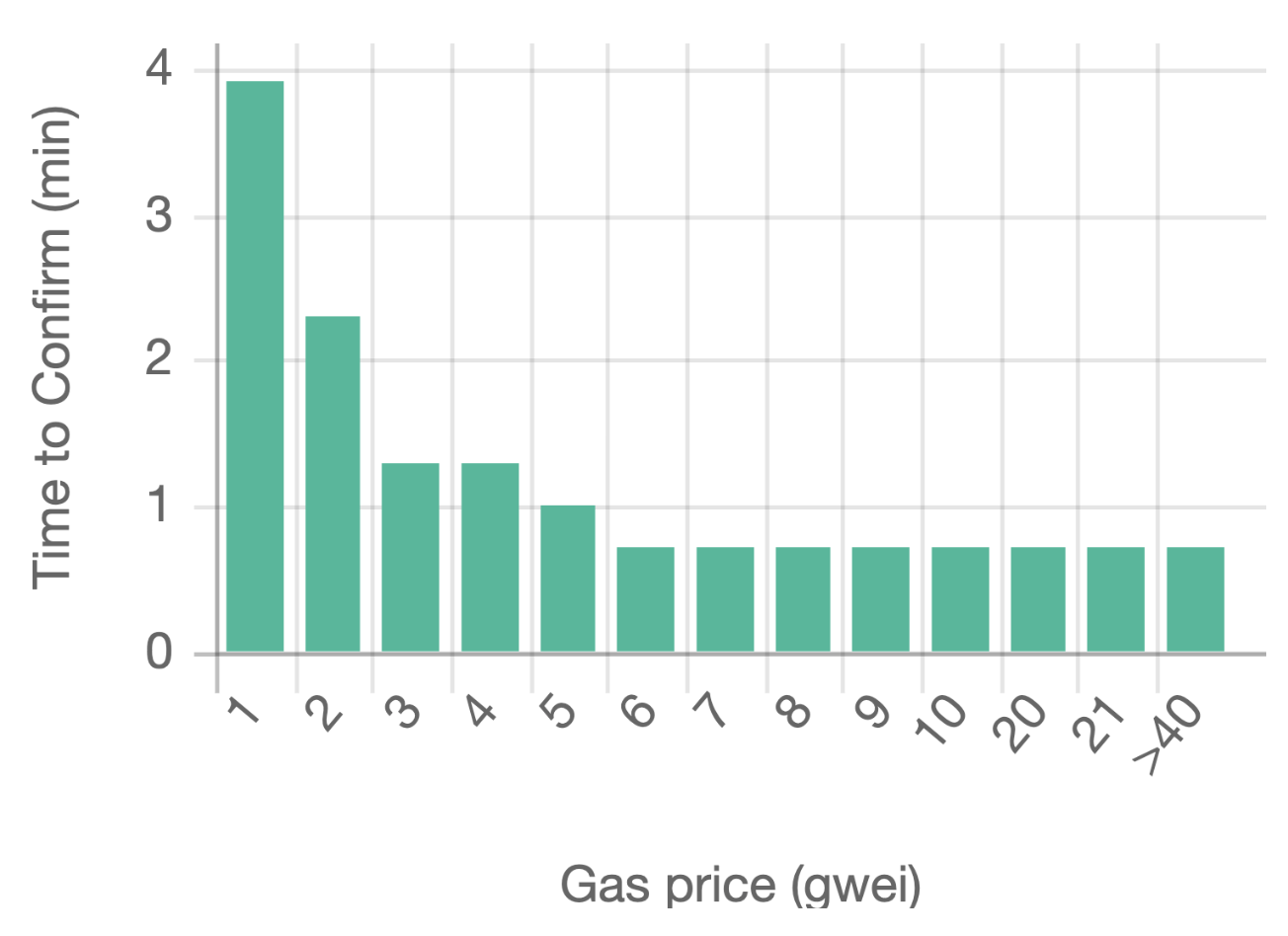}
  \caption{Transaction Confirmation Time versus Gas Price, Source data: Ethereum Gas Station \cite{ethgasstation}
}
  \label{fig:graph-confirmation-time}
\end{figure}

The Ethereum Gas Station \cite{ethgasstation} publishes live statistics on how quickly transactions will be processed based on the gas price specified for a transaction. For example, on February 18, 2019, it showed that some miners would process transactions at 1.0 gwei, however only twenty-six blocks in the past 200 had included transactions at this price. If a user was prepared to pay 6.0 gwei, then their transaction was likely to be processed within the next two blocks. Given an average block time of 15 seconds, this translates to transactions possibly being processed sometime in the next fifty minutes for gas prices of 1.0 gwei or with a high degree of probability processed in the next thirty seconds for gas prices of 6.0 gwei. This range of confirmation time versus gas price is shown in Figure  \ref{fig:graph-confirmation-time}. 
Confirmation times for gas prices below 1.0 gwei are either very high, or the transaction does not get mined at all. As the gas price increases, the confirmation time decreases. At 6.0 gwei, the transaction is likely to go into the next block. As such, there is no benefit to users for offering to pay gas prices above 6.0 gwei as the transaction is already likely to be mined as soon as it can be. 

Gas pricing in Ethereum is extremely fluid. This 6.0 gwei value is double the value for which transactions could have been mined for expeditiously in August 2018 \cite{robinson2018b}. This increase is due to the current high network load which is seeing the Ethereum blocks at 100\% capacity. However, at the same time the gas price required to execute a transaction expeditiously has doubled, the price of Ether has approximately halved. As such, the US\$ price of executing Smart Contracts is approximately the same now in February 2019 as it was in August 2018.

The gas usage for the reference implementation of the Sidechain Anonymous Pinning contract is shown in Table \ref{table_gas}. The values reflect the fact that the base cost of all transactions in Ethereum is 21,000 gas, that each write to a new storage location costs 20,000 gas, that subsequent writes to storage locations costs 5,000 gas, that gas is refunded for deleting storage locations, and that most other common functions cost relatively little compared to the cost of data storage. In the table, the gas cost to US\$ conversion is calculated based on an Ether price of US\$148, and the fact that 1 Ether is 10\textsuperscript{9} gwei. 
For transactions which are not time critical and involve a lot of gas, such as contract deployment, it makes sense to specify a low gas price such as 1.0 gwei. At this price, the Sidechain Anonymous Pinning contract could be deployed for US\$0.371. For transactions which are time critical such as voting or pinning, a higher gas cost should be paid to ensure the transaction is mined quickly. Adding a Pin, with a gas price of 6.0 gwei, would cost US\$0.058. 

Pinning strategies are likely to be dependant on the deployment scenario. For some deployments, there may be transactions irregularly, say once per month. In this scenario, pinning each block which contains a transaction makes sense. In other scenarios, Pins might be posted once every one hundred transactions, with at least one Pin posted per day, but not more than one Pin per five minutes. In yet other scenarios, Pins might be posted each ten blocks. As such, determining the cost of using this technology is situation dependant. However, assuming one Pin per hour, this would add up to an annual cost of US\$508.

\begin{table*}[t]
  \centering
  \begin{threeparttable}
  \begin{tabular}{| l | l | r | r |}
    \hline
    Function Call                                    & Scenario                                                           & Gas Used   & US\$*    \\
    \hline
    \hline
    SidechainAnonPinningV1                & Contract deployment                                        &   2562653   & 0.379   \\
    \hline
    addSidechain                                   & Add a sidechain to be managed by the contract &   133242   & 0.118   \\
    \hline
    addPin                                              & Add a pin to the pin map                                 &   64972   & 0.058   \\
    \hline
    \multirow{5}{*}{proposeVote}             & Propose to add an unmasked participant           &   151007   & 0.134   \\ 
                                                               \cline{2-4}
                                                             & Propose to add a masked participant             &   121635   & 0.108   \\
                                                               \cline{2-4}
                                                             & Propose to remove an unmasked participant &   136646   & 0.121   \\
                                                               \cline{2-4}
                                                             & Propose to remove a masked participant       &   122234   & 0.109   \\
                                                               \cline{2-4}
                                                             & Propose to contest a pin                                 &   156481   & 0.139   \\
    \hline
    vote                                                  & Vote for or against a proposed vote. Same gas used for all scenarios   &   54458   & 0.048   \\
    \hline
    \multirow{5}{*}{actionVotes}              & Action a successful vote to add an unmasked participant &   67663  & 0.060   \\
                                                               \cline{2-4}
                                                              & Action a successful vote to add a masked participant &   72626   & 0.064   \\
                                                               \cline{2-4}
                                                              & Action a successful vote to remove a masked participant         &   47148   & 0.042   \\
                                                               \cline{2-4}
                                                              & Action a successful vote to remove an unmasked participant   &   38890   & 0.035   \\
                                                               \cline{2-4}
                                                              & Action a successful vote to contest a pin                                   &   38891   & 0.035   \\
    \hline
    \hline
  \end{tabular}
   \begin{tablenotes}
      \small
      \item *Assumes Ether price of US\$148, 1.0 gwei for contract deployment and 6.0 gwei for other transactions.
    \end{tablenotes}
  \end{threeparttable}
  \caption{Gas Cost Estimates}
  \label{table_gas}
\end{table*}

\section{Security Proof}
To show that anonymous pinning is secure we need to demonstrate that knowledge of two successive MapKey values, which may be obtained by observing a Pin being challenged, cannot be used to compute earlier MapKey values. Attackers who can determine which MapKey - Pin combinations belong to which private blockchain can determine the pinning rate of the blockchain and may be able to infer the transaction rate of the blockchain. Additionally, the proof must demonstrate that attackers after a successful challenge to a Pin, can not determine the replacement MapKey value MapKey\textsubscript{t}\textsuperscript{*}. This is important as an attacker who can calculate values of MapKey\textsubscript{t}\textsuperscript{*} could maliciously insert a invalid Pin values into the map at MapKey\textsubscript{t}\textsuperscript{*} addresses, thus preventing participants from posting any valid Pins.

\subsection{Assumptions}
We assume that efficient computation of preimages for the function \texttt{KECCAK-256()} is infeasible. That is, that given a hash value \texttt{h}, it is infeasible to find a value \texttt{x} such that \texttt{KECCAK-256(x) = h}.

We also assume that the PRF is secure in both forward and reverse directions, without knowledge of the secret seed value, the Private Blockchain Secret. In other words, that functions \texttt{F\textsubscript{PRF+}} and \texttt{F\textsubscript{PRF-}}  such that \texttt{F\textsubscript{PRF+}(PRF(t)) = PRF(t+1)} and \texttt{F\textsubscript{PRF-}(PRF(t) = PRF(t-1))} cannot be implemented without knowledge of the PRF secret seed value.  
 
We assume that an attacker can observe all activity on the pinning Smart Contract. As such, the attacker knows all MapKey, Pin pairs which have been submitted to the contract. Further, the attacker knows all Private Blockchain Identifier values which are used in the Smart Contract. 

\subsection{Reverse Proof}
The attacker, by observing a contested Pin value, corresponding to time \texttt{t}, obtains the following: 

\begin{itemize}
\item MapKey\textsubscript{t-1}, MapKey\textsubscript{t}
\item PRF(t)
\item Pin\textsubscript{t-1}
\item PBI (assumed to be known by all parties)
\end{itemize}

As earlier defined: \texttt{MapKey\textsubscript{t} = KECCAK-256( PBI, Pin\textsubscript{t-1}, PRF(t))}. The attacker wishes to compute \texttt{F\textsubscript{REV}(MapKey\textsubscript{t-1}, MapKey\textsubscript{t}, PRF(t), all existing (MapKey, Pin) combinations) = MapKey\textsubscript{t-2}}. 

Since the previous MapKey value must be derived from the previous Pin and the previous PRF value, any possible \texttt{F\textsubscript{REV}} must determine the previous Pin from the Pin values previously submitted to the contract, and must compute PRF(t-1) from either MapKey\textsubscript{t} or PRF(t). This implies the existence of either a preimage attack against \texttt{KECCAK-256}, the existence of \texttt{F\textsubscript{PRF-}}, or both, which violates our assumptions.

Therefore, subject to the assumptions above, \texttt{F\textsubscript{REV}} does not exist.

\subsection{Forward Proof}
The attacker has available the same information available as in the Reverse Proof. The attacker wishes to compute  \texttt{F\textsubscript{FWD}(MapKey\textsubscript{t-1}, MapKey\textsubscript{t}, PRF(t)) = MapKey\textsubscript{t}\textsuperscript{*}}.

Since the MapKey\textsubscript{t}\textsuperscript{*} value must be derived from the next PRF value, any possible  \texttt{F\textsubscript{FWD}} must compute PRF(t+1) from PRF(t). This implies the existence of \texttt{F\textsubscript{PRF+}} which violates our assumptions.

Therefore, subject to the assumptions above, \texttt{F\textsubscript{FWD}} does not exist.

\section{Keyed Hash Usage of Keccak }
KECCAK-256 \cite{keccak} is used as a keyed hash when calculating MapKey\textsubscript{t}. As KECCAK is not subject to length extension attacks, this usage is not vulnerable to such attacks.

\section{Affects of Pinning on Finality}
When a private blockchain is pinned to a Management Chain, a block may only be considered final when a Pin representing the block can no longer be challenged. If the Management Chain is Ethereum MainNet, then Ethereum MainNet's probabilistic finality needs to be considered. Values are often read back from Ethereum MainNet six blocks from the end of the chain to ensure chain reorganisations are extremely unlikely to change values read from the distributed ledger. Blocks closer to the end of the chain can be used. However there is a risk that the block will be removed, and hence a different value returned. As such, by the time observers see a Pin being posted that they would wish to contest, the Pin could be six blocks old. 

If the participant which wishes to contest a Pin is a Masked Participant, they would need to submit a transaction to unmask themselves prior to commencing the process to challenge the Pin. They then need to submit a transaction to the \texttt{proposeVote} function, proposing that the Pin be contested. Other participants which see the false Pin are likely to simultaneously submit transactions to the \texttt{proposeVote} function. Only one transaction will successfully start the voting process and the others will be rejected. Six blocks after the start of the \texttt{proposeVote} function, all participants on the private blockchain will be able to see the vote is active and vote. Once the voting period expires, but before the Pin Contest Period expires, the vote can be actioned using the \texttt{actionVotes} function. This means, when pinning to Ethereum MainNet, the Voting Period should be at least seven blocks, and the Pin Contest Period should be at least fifteen blocks: six blocks to see the Pin, plus one block to unmask, plus a Voting Period of seven blocks, plus one block to action the vote. Given an Ethereum MainNet block period of 15 seconds, this is 225 seconds or almost four minutes.

If the Management Chain is another private blockchain which uses IBFT with instant finality, then the Pin Contest Period could be reduced to five blocks: one block to see the Pin, plus one block to unmask, plus a Voting Period of two blocks, plus one block to action the vote. As the block period of a private blockchain is implementation specific, there is no was to relate the number of blocks on the private blockchain to time, for all blockchains.

The period of time between when Pins of a private chain are posted, the Pinning Period, need to be considered. Though it could be possible to post a Pin for each block, it may be more desirable to pin once each time period, say each five minutes, each hour, or each day. By pinning only once per Pinning Period, it means that the finality of a block created at the start of the Pinning Period would be equal to Pinning Period plus Pin Contest Period. 

If a hierarchical pinning approach is used, then for each pinning layer there is an extra Pin Contest Period and Pinning Period to be considered. The Pin Contest Period and Pinning Period for each layer should be summed to give the worst case private blockchain block finality time.

\section{Conclusion}
Private Blockchains which have their state pinned using the Anonymous State Pinning approach described in this paper can protect the blockchain from the majority of participants colluding to revert the state of the blockchain, while maintaining the privacy of participants and hiding the rate that pins of the state are posted to a Management Chain. An analysis of an implementation of the approach has shown that it would cost US\$508 per year to pin the state of a private blockchain to Ethereum MainNet, assuming a pinning rate of one Pin per hour, and current Ether and Gas prices. A security analysis has shown that the technique does not reveal information which could be used to determine future and previous MapKey - Pin values, or the pinning rate of the private blockchain. The analysis of block finality has shown that if the private blockchain was pinned to Ethereum MainNet, that blocks would not be considered final until a Pin Contest Period of four minutes has expired plus the period of time between when Pins are posted to Ethereum MainNet, the Pinning Period.

\ifCLASSOPTIONcompsoc
  \section*{Acknowledgments}
\else
  \section*{Acknowledgment}
\fi
This research has been undertaken whilst we have been employed full-time at ConsenSys. We acknowledge Dr Catherine Jones, Dr Marius Portmann, and Dr Sandra Johnson for their thoughtful review comments and suggestions.



\bibliographystyle{IEEEtran}
\bibliography{IEEEabrv,ref}
%
%
%

\end{document}